\documentclass[pdftex,twocolumn,epjc3]{svjour3}          

\RequirePackage[T1]{fontenc}
\usepackage{amsmath}
\usepackage{amssymb}
\usepackage{lipsum}
\smartqed  
\usepackage{graphicx}
\RequirePackage{graphicx}
\RequirePackage{mathptmx}      
\RequirePackage{flushend}
\RequirePackage[numbers,sort&compress]{natbib}
\RequirePackage[colorlinks,citecolor=blue,urlcolor=blue,linkcolor=blue]{hyperref}

\journalname{Eur. Phys. J. C}
\usepackage[greek, english]{babel}
\usepackage{alphabeta}
\usepackage{textgreek}
\def\lcdm{\textLambda CDM }
\def\mcT{\mathcal{T}}
\begin{document}

\title{The cosmology of $f(R, L_m)$ gravity: constraining the background and perturbed dynamics}


\author{Shambel Sahlu\thanksref{e1,addr3,addr31,addr32}
\and
Alnadhief H. A. Alfedeel\thanksref{e2,addr1,addr2,addr3} 
        \and
        Amare Abebe \thanksref{e3,addr3,addr4}
}

\thankstext{e1}{e-mail: shambel.sahlu@nithecs.ac.za}
\thankstext{e2}{e-mail: aaalnadhief@imamu.edu.sa}
\thankstext{e3}{e-mail: amare.abebe@nithecs.ac.za}
\institute{Centre for Space Research, North-West University, Potchefstroom 2520, South Africa\label{addr3}
\and
Department of Physics, Wolkite University, Wolkite, Ethiopia\label{addr31}
\and 
Entoto Observatory and Research Center, Space Science and Geospatial Institute, Ethiopia \label{addr32}
\and
Department of Mathematics and Statistics, Imam Mohammad Ibn Saud Islamic University (IMSIU),Riyadh\label{addr1}
          \and
          Department of Physics, Faculty of Science, University of Khartoum, Sudan \label{addr2}
              \and
              National Institute for Theoretical and Computations Sciences (NITheCS), South Africa\label{addr4}
}

\date{Received: date / Accepted: date}

\maketitle
\begin{abstract}

This paper delves into the late-time accelerated expansion of the universe and the evolution of cosmic structures within the context of a specific \( f(R, L_m) \) gravity model, formulated as \( f(R, L_m) = \lambda R + \beta L_m^\alpha + \eta \). To study the cosmological viability of the model,  we employed the latest cosmic measurement datasets: i) 57 observational Hubble parameter data points (\texttt{OHD}); ii) 1048 distance moduli data points (\texttt{SNIa}); iii) a combined dataset (\texttt{OHD+SNIa}); and large scale structure datasets, including iv) 14 growth rate data points (\texttt{f}); and v) 30 redshift space distortion data points (\texttt{f}$\sigma_8$). These datasets facilitated the constraint of the \( f(R, L_m) \)-gravity model via MCMC simulations, followed by a comparative analysis with the \(\Lambda\)CDM model. A comprehensive statistical analysis has been conducted to evaluate the \( f(R, L_m) \)-gravity model's efficacy in explaining both the accelerated expansion of the universe and the growth of cosmic structures. Using large-scale structure data, we find the best-fit values of $\Omega_m = 0.242^{+0.016}_{-0.032}$, $\alpha = 1.15^{+0.20}_{-0.20}$, $\beta = 1.12^{+0.13}_{-0.30}$, $\lambda = 0.72^{+0.30}_{-0.13}$  and \(\gamma = 0.555\pm {0.014}\)  using \texttt{f}-data and $\Omega_m = 0.284^{+0.035}_{-0.049}$, $\sigma_8 = 0.799^{+0.045}_{-0.086}$, $\alpha = 0.766^{+0.026}_{-0.064}$, $\beta = 1.08^{+0.40}_{-0.16}$, and $\lambda = 0.279^{+0.078}_{-0.11}$ using \texttt{f}\(\sigma_8\)-data at the $1\sigma$ and $2\sigma$ confidence levels, respectively, with the model showing substantial observational support based on $\Delta$AIC values but less observational support based on the $\Delta$BIC values on Jeffreys' statistical criteria. On the other hand, from the joint analysis of the \texttt{OHD+SNIa} data, we obtain $\alpha = 1.091^{+0.035}_{-0.042}$, \(\beta = 1.237^{+ 0.056}_{-0.16}\) and $\lambda = 0.630^{+0.031}_{-0.050}$ with the Jeffreys' scale statistical criteria showing the \( f(R, L_m) \) model having substantial support when using \texttt{OHD} data, less observational support with the joint analysis \texttt{OHD+SNIa}, and rejected using \texttt{SNIa} data, compared with $\Lambda$CDM at the background level.
\end{abstract}
\section{Introduction}
Over the last couple of decades, several observational data \cite{perlmutter1998discovery,riess1998observational,scolnic2018complete}
have consistently demonstrated that the universe is currently undergoing an accelerating expansion. The acceleration is generally assumed to have been caused by the enigmatic
dark energy (DE), which is formally represented by the cosmological constant $\Lambda$ in Einstein's gravitational field equations. Despite its nature being unknown, it is regarded as one of, if not the most significant, cosmological challenges to date. Several hypotheses have been devised to address the issue of dark energy and dark matter (DM), another unknown component of the matter-energy content whose existence is inferred through its gravitational effects.
There are several avenues to addressing these challenges: e.g., we can consider time-varying (dynamical) DE models or introducing exotic matter fields,  which are currently beyond the scope of this article, or we can modify the general theory of relativity (GR). The primary goal of the modified theory of gravity (MTG) is to forecast the cosmic acceleration that occurs at later times by modifications to the Einstein's-Hilbert action of GR.
For
instance, the modification of Einstein's equations for the gravitational field includes introducing the functions $f(R)$ \cite{buchdahl1970non,nojiri2011unified} in the Einstein-Hilbert action as opposed to just the Ricci scalar $R$ in standard general relativity (GR).  $f(T)$ gravity \cite{bengochea2009dark,linder2010einstein} is another modification of gravity that incorporates torsion within the framework of the Teleparallel Equivalent of General Relativity (TEGR). The concept of $f(Q)$ gravity \cite{jimenez2018coincident,beltran2019geometrical} attributes gravity to the nonmetricity $Q$ of the metric, which mathematically specifies how the length of a vector changes during parallel transport. The approach is referred to as symmetric teleparallel gravity (STEGR). Other more exotic models considered in the literature include $f(R, \mcT)$ \cite{harko2011f} and $f(Q,\mcT)$ \cite{xu2019f} where $\mcT$ here denotes the trace of the energy-momentum tensor.
\\
Another addition to an already crowded field of modifications to gravity is the $f(R,L_m)$ model proposed by  \cite{RL,jaybhaye2023constraining}, $L_m$ denoting the matter Lagrangian density. 
Being a relative newcomer, the model promises to shed light on the coupling between matter and geometry and the cosmological implications thereof, but there has so far been no comprehensive work in terms of constraining data coming from the background evolution and large-scale structure of the universe. Most of the existing work has focused on constraints from the background evolution (see, for example, the work in \cite{mkaa} and references therein) but even then the focus has been on very specific forms of the coupling.
\\
\\
Motivated by the above discussion, in this manuscript, we constrain the cosmological model's free parameters in the context of a more general model chosen to have a valid \lcdm limit, $f(R, L_m) = \lambda R + \beta L_m^\alpha + \eta$, using the observational data set of the background expansion history, and then investigate the cosmological perturbations and the growth of large-scale structure. Moreover, we compare the model against the traditional $\Lambda$CDM model and test the validity of the model using well-known statistical criteria.
\\
\\
The paper is organized in the following manner: In Section~\ref{efes}, we introduce the alternative theory of gravity $f(R, L_m)$. In this section, the modified field equations within the FLRW geometry are outlined. In Section~\ref{datameth}, we present the data and methodology. The free parameters of the cosmological model, represented by $f(R,L_m) = \lambda R + \beta L_m^\alpha + \eta $, are determined by constraining them using the observational data sets discussed in Section~\ref{model}. 
The perturbations and the structure growth parameters in the framework of the $f(R, L_m)$ gravitation are explained in Sections~\ref{pert} and \ref{growth}. The results and consequences of the article are reported in Section~\ref{conc}.
\section{The $f(R,L_m)$ theory of gravity} \label{efes}
The generalized form of the Einstein-Hilbert action for the modified theory of gravity $f(R,L_m)$ that was introduced in \cite{RL, nojiri2004gravity,jaybhaye2022cosmology,jaybhaye2023baryogenesis,allemandi2005dark} is given by the following expression:
\begin{equation}\label{1}
S= \int{f(R,L_m)\sqrt{-g}d^4x}, 
\end{equation}
where $f(R,L_m)$ is an arbitrary function of the Ricci scalar $R$ and $L_m$, the matter Lagrangian term $L_m$. Here $g$ is the determinant of the metric tensor. Throughout this paper natural units $8 \pi G=c=1$ will be adopted unless stated otherwise.
The field equations are obtained by varying the action in Eq.\eqref{1} with respect to the metric tensor $g_{\mu\nu}$ as follows:
\begin{equation} \label{2}
f_{R} R_{\mu \nu} + \left( g_{\mu \nu} \square - \nabla_{\mu} \nabla_{\nu} \right) f_R - \frac{1}{2} \left( f - f_{L_m} L_m  \right) g_{\mu \nu} = \frac{1}{2} f_{L_m} T_{\mu\nu}~\;,
\end{equation}
where $f_R  = \partial f/ \partial R $, $f_{L_m} = \partial f / \partial L_m$ and $T_{\mu\nu}$ is the energy-momentum tensor of the matter content ``a perfect type fluid'', which is defined by
\begin{equation}\label{3}
T_{\mu\nu} = -\frac{2}{\sqrt{-g}} \frac{\delta(\sqrt{-g}L_m)}{\delta g^{\mu\nu}}~.
\end{equation}
The general relativity (GR) limit can be recovered if $f(R,L_m) =R/2+L_m$, obtained when $\alpha=1=\beta$ and $\lambda=1/2,\eta=0$. The \lcdm limit, on the other hand, requires setting $\eta=-\Lambda$. By contracting Eq. \eqref{2}, we could obtain an equation of a similar type that is dependent on the trace of the energy-momentum tensor ${\mathcal{T}}$, the Ricci scalar $R$, and the Lagrangian density of matter $L_m$ as
\begin{equation}\label{4}
R f_R + 3\square f_R - 2(f-f_{L_m}L_m) = \frac{1}{2} f_{L_m} T~,
\end{equation}
where $\square F$ is the d'Alembertian of a scalar function $F$, given by 
\[\square F = \frac{1}{\sqrt{-g}} \partial_\alpha (\sqrt{-g} g^{\alpha\beta} \partial_\beta F)~.\]
Taking the covariant derivative of Eq. \eqref{2}, produced auxiliary equation as 
\begin{equation}\label{5}
\nabla^\mu T_{\mu\nu} 
= 2\nabla^\mu ln(f_{L_m}) \frac{\partial L_m}{\partial g^{\mu\nu} } = \nabla^\mu ln(f_{L_m}) \{ T_{\mu \nu} - g_{\mu \nu} L_m \}\;.
\end{equation}
This equation determines the evolution equation for the energy density of the specific fluid you have selected.
We consider the spatially homogeneous and isotropic Universe of a flat Friedmann-Lema\^{i}tre-Robertson-Walker (FLRW) metric,
\begin{equation}\label{9}
ds^2= -dt^2 + a^2(t)[dx^2+dy^2+dz^2]~,
\end{equation}
where $a(t)$ is the universe scale factor that measures the expansion
of the universe at a certain time $t$. In the context of this metric, the Ricci scalar is calculated as
\begin{equation}\label{12}
R= 6 ( \dot{H}+2H^2 )~,
\end{equation}
where $H\equiv\dot{a}/a$, $\dot{a}$ are the Hubble parameter and the time derivative of the scale factor $a$ respectively. 

For a universe that is filled by  a perfect fluid, the energy-momentum tensor is given by 
the following expression: 
\begin{equation}\label{13}
T_{\mu\nu} = (\rho+p) u_{\mu} u_{\nu} + p g_{\mu\nu}\;,
\end{equation}
where $\rho$, $p$, and $u^\mu=(1,0,0,0)$ are the matter-energy density, the spatially isotropic pressure, and the fluid's four-velocity vector, respectively. 

Within the context of $f(R, L_m)$ gravity, the generalized Friedmann equations that govern the Universe's dynamics can be written as:
\begin{eqnarray}\label{14}
&&3H^2 f_R + \frac{1}{2} \left( f-f_R R-f_{L_m}L_m \right) + 3H \dot{f_R}= \frac{1}{2}f_{L_m} \rho~,\\
&&\label{15}
\dot{H}f_R + 3H^2 f_R - \ddot{f_R} -3H\dot{f_R} + \frac{1}{2} \left( f_{L_m}L_m - f \right) = \frac{1}{2} f_{L_m}p~.
\end{eqnarray}  
\section{Data and methodology}\label{datameth}
In this manuscript, we consider the recent cosmic measurement data namely: \texttt{OHD and SNIa};  and the growth measurement data: including the growth rate \texttt{ f}  and the redshift space distortion \texttt{f}$\sigma_8$ for further analysis. 
\begin{itemize}
\item {Type I Supernova data}. We use Type I Supernova distance moduli measurements from the Pantheon \cite{scolnic2018complete}, which consists of 1048 distinct SNeIa ranging in the redshift interval $z \in [0.001, 2.26]$. We refer to this dataset as \texttt{SNIa}. 

\item {Observational Hubble parameter data}: 
We consider the measurements of the expansion rate Hubble parameter $H(z)$ which consists of 57 data points in total \cite{yu2018hubble, chen2017determining,yadav2021constraining, dixit2023observational},  31 data points from the relative ages of massive, early-time, passively-evolving galaxies, known as cosmic chronometers (CC) with 26 data points
the baryon acoustic oscillations (BAO), which are provided by the Sloan Digital Sky Survey (SDSS), DR9, and DR11   $0.0708 < z \leq 2.36$. We refer to this dataset as \texttt{OHD}. 
\item {Large scale structure data:}  We have implemented the sets of redshift-space distortions data \texttt{f}$\sigma_8$ with the latest separate measurements of the growth rate \texttt{f}-data from the VIPERS and SDSS collaborations, (see Table \ref{datasetsofgrowth}).
In particular, we will use:  
$i)$  the  30  redshift-distortion measurements of \texttt{f}$\sigma_8$, dubbed \texttt{f}$\sigma_8$, covering the redshift range $0.001\leq z\leq 1.944$ and; $ii)$ fourteen  \texttt{f}$\sigma_8$ data points in the red-shift ranges of $0.001\leq z\leq 1.4$.
\item Different Software and Python packages  MCMC hammer \cite{foreman2013emcee}, GetDist \cite{lewis2019getdist} are considered to parameter estimations.
\begin{table}[h!]
{\small

\centering
\begin{tabular}{|c c c c|} 
 \hline
 Dataset & $z$ &  $f$  &Ref.\\ 
 \hline
 ALFALFA &0.013 &   0.56$\pm$0.07& \cite{avila2021growth} \\
 SDSS & 0.10 &  0.464 $\pm$ 0.040&  \cite{shi2019mapping}\\
 SDSS-MGS &0.15&0.490$\pm$ 0.145&\cite{HowlettC}\\
 GAMA &0.18&0.49$\pm$0.12&\cite{BlakeC1}\\
 Wigglez &0.22 &0.6$\pm$0.10& \cite{blake2011wigglez}\\
SDSS-LRG &0.35 &    0.7$\pm$0.18&\cite{tegmark2006cosmological}\\
WiggleZ&0.41 &    0.7$\pm$0.07&\cite{blake2011wigglez}\\
2SLAQ &0.55 &    0.75$\pm$0.18&\cite{ross20072df}\\
 WiggleZ&0.60& 0.73$\pm$0.07&\cite{BlakeC}\\
WiggleZ&0.77 &    0.91$\pm$0.36&\cite{guzzo2008test}\\
 Vipers PDR-2&0.60&0.93 $\pm$ 0.22& \cite{De11,shambel}\\
 Vipers PDR-2&0.86&0.99 $\pm$ 0.19&\cite{De11,shambel}\\
 VIMOS-VLT DS &0.77&0.91 $\pm$ 0.36&\cite{wang2018clustering}\\
 2QZ\&2SLAQ&1.40 &    0.90$\pm$0.24&\cite{daangela20082df}\\
 
 \hline
 Dataset & $z$ & $f\sigma_8$  &Ref.\\ 
 \hline
 2MTF & 0.001 & 0.505 $\pm$ 0.085 & \cite{Howlett}\\ 
 6dFGS+SNIa & 0.02& 0.428 $\pm$ 0.0465 &\cite{HutererD} \\
 IRAS+SNIa & 0.02 & 0.398 $\pm$ 0.065 &  \cite{HudsonMJ,TurnbullSJ}\\
 2MASS & 0.02 & 0.314 $\pm$ 0.048 & \cite{HudsonMJ,Davis}\\
 SDSS & 0.10 & 0.376 $\pm$ 0.038 &  \cite{shi2019mapping}\\
 SDSS-MGS &0.15&0.490$\pm$ 0.145&\cite{HowlettC}\\
 2dFGRS &0.17&0.510 $\pm$ 0.060&\cite{SongYS}\\
 GAMA &0.18&0.360 $\pm$ 0.090&\cite{BlakeC1}\\
 GAMA&0.38&0.440 $\pm$ 0.060& \cite{BlakeC1}\\
 SDSS-LRG-200&0.25&0.3512 $\pm$ 0.0583&\cite{SamushiaL}\\
 SDSS-LRG-200&0.37&0.4602 $\pm$ 0.0378&\cite{SamushiaL}\\
 BOSS DR12&0.31&0.469 $\pm$ 0.098&\cite{YWang}\\
 BOSS DR12&0.36&0.474 $\pm$ 0.097&\cite{YWang}\\
 BOSS DR12&0.40&0.473 $\pm$ 0.086&\cite{YWang}\\
 BOSS DR12&0.44&0.481 $\pm$ 0.076&\cite{YWang}\\
 BOSS DR12&0.48&0.482 $\pm$ 0.067&\cite{YWang}\\
 BOSS DR12&0.52&0.488 $\pm$ 0.065&\cite{YWang}\\
 BOSS DR12&0.56&0.482 $\pm$ 0.067&\cite{YWang}\\
 BOSS DR12&0.59&0.481 $\pm$ 0.066&\cite{YWang}\\
 BOSS DR12&0.64&0.486 $\pm$ 0.070&\cite{YWang}\\
 WiggleZ&0.44&0.413 $\pm$ 0.080&\cite{BlakeC}\\
 WiggleZ&0.60&0.390 $\pm$ 0.063& \cite{BlakeC}\\
 WiggleZ&0.73&0.437 $\pm$ 0.072&\cite{BlakeC}\\
 Vipers PDR-2&0.60&0.550 $\pm$ 0.120&\cite{De11,shambel}\\
 Vipers PDR-2&0.86&0.400 $\pm$ 0.110&\cite{De11,shambel}\\
 FastSound&1.40&0.482 $\pm$ 0.116&\cite{okumura2016subaru}\\
 SDSS-IV&0.978&0.379 $\pm$ 0.176&\cite{wang2018clustering} \\
 SDSS-IV&1.23&0.385 $\pm$ 0.099&\cite{wang2018clustering}\\
 SDSS-IV&1.526&0.342 $\pm$ 0.070&\cite{wang2018clustering}\\
 
 SDSS-IV&1.944&0.364 $\pm$ 0.106&\cite{wang2018clustering}\\
 \hline
\end{tabular}
\caption{This study incorporates a collection of structure data, consisting of 30 data points for redshift space distortion ($f\sigma_8$), and fourteen data points for the growth rate ($f$).
}
 \label{datasetsofgrowth}
 }
\end{table}
\end{itemize}
\section{Background cosmology in $f(R,L_m)$ gravity}\label{model}
To confront the model with observational data, we consider the following general functional form of $f(R,L_m)$:
\begin{equation}\label{16} 
f(R,L_m)= \lambda R + \beta L_m^\alpha +\eta ~,
\end{equation}
where $\alpha, \beta, \lambda$, and $\eta$ are constants that will be determined from the cosmological observational data as we will see later.  
When $\alpha=1, \beta=1$ and $\lambda=1/2$, the classic Friedmann equations of GR are retrieved, specifically.
When plugging Eq.\eqref{16} into Eqs. \eqref{14} and \eqref{15} yields: 
\begin{eqnarray}\label{17}
&&3H^2= \frac{\beta}{2\lambda} \left[ \alpha  L_m^{\alpha-1} \rho - (1-\alpha) L_m^{\alpha} \right] - \frac{\eta}{2\lambda}~,\\
&&2 \dot{H}+ 3H^2 = \frac{\beta}{2\lambda} \left[ (\alpha-1) L_m^{\alpha} - \alpha p  L_m^{\alpha-1} \right] - \frac{\eta}{2\lambda}~.\label{18}
\end{eqnarray}
These equations, as we may see, entirely depend on the matter Lagrangian $L_m$ form choice and the form of the matter Lagrangian, and the energy-
momentum tensor, are strongly dependent on the equation
of state (EoS) \cite{harko2014generalized}.

As indicated in \cite{harko2015gravitational, harko2014generalized}, we shall now proceed using the natural and simple form of $L_m=\rho$, which corresponds to a dust-fluid particle.
Therefore, substituting $L_m=\rho$ into Eqs. \eqref{17} and \eqref{18}, the generalized Friedman
equations may be reformulated as:  
\begin{eqnarray}\label{17a}
&&3H^2= \frac{\beta}{2\lambda}(2\alpha-1) \rho^{\alpha} - \frac{\eta}{2\lambda} ~,\\
&&\label{18a}
2 \dot{H}+ 3H^2= \frac{\beta}{2\lambda}\left[(\alpha-1)\rho - \alpha p \right]\rho^{\alpha-1} - \frac{\eta}{2\lambda} ~.
\end{eqnarray}
The divergence of the energy-momentum tensor $T_{\mu \nu}$ gives 
\begin{equation} \label{19}
  (2\alpha -1) \dot{\rho}_m + 3 \Gamma H \rho_m= 0~,  
\end{equation} 
which can be integrated with respect to cosmic time $t$ to produce
\begin{equation} \label{20}
  \rho = \rho^\alpha_0 a^{\frac{-3 \Gamma}{ (2\alpha -1)}}~,  
\end{equation}
where $\rho_0$ is the current energy density, $\gamma =w+1$, with $w$ being equation of state parameter of the cosmic fluid.
Inserting the expression of $\rho$ into Eq. \eqref{17} and rearrange for $H$ yields
%
\begin{equation}\label{22}
H(z)=  H_0\sqrt{ \frac{(2\alpha-1)\beta}{2\lambda} \Omega_{m} (1+z)^{ \frac{3\alpha \gamma}{(2\alpha -1)} } - \frac{\Omega_{\eta}}{2\lambda}} ~,    
\end{equation}
where $\rho_0^\alpha  = 3 H_0^2 \Omega_{m}$ , $\eta= 3H_0^2 \Omega_{\eta}$. $\Omega_m$ and  $\Omega_{\eta}$ referees to the normalized energy density of matter fluid and dark energy. From Eq. \eqref{22}, the normalized energy density $E(z)\equiv H(z)/H_0$ define as  
\begin{equation}\label{22a}
E(z)= \sqrt{ \frac{(2\alpha-1)\beta}{2\lambda} \Omega_{m0} \left[ (1+z)^{ \frac{3\alpha \gamma}{(2\alpha -1)} } -1 \right] +1 }\;,  
\end{equation}
where $\Omega_{\eta} = (2\alpha-1)\beta \Omega_{m} -2\lambda$.   For the case of $\alpha = \beta =1$ and $\lambda=1/2$, $\Omega_{\eta}$ yields as $ \Omega_{\eta} = -1+\Omega_m = -\Omega_\Lambda$ and Eq. \eqref{22a} reduced to $\Lambda$CDM limit as
\begin{eqnarray}
    E(z)= \sqrt{\Omega_{m}(1+z)^3+1-\Omega_m}\;.
\end{eqnarray}
 Combining Eqs. \eqref{17a} and \eqref{18a}, the the deceleration parameter for the case of $f(R,L_m)$  it becomes
\begin{eqnarray}
 && q(z) =   -1 \nonumber\\&& + \frac{3 \alpha \beta \gamma}{4\lambda} \left[\frac{ \Omega_{m0} (1+z)^{ \frac{3\alpha \gamma}{2\alpha-1} } }{  \frac{(2\alpha-1)\beta \Omega_{m0}}{2\lambda} \left( (1+z)^{ \frac{3\alpha \gamma}{(2\alpha -1)} }- 1 \right) +1 }\right]
  \;,
\end{eqnarray}
and the effective equation of state EoS which relates the cosmological fluid pressure to its energy density is calculated as follows:
\begin{equation}\label{EoS}
    w_{eff}(z) =  -\frac{1}{3}+\frac{2q(z)}{3}\;.
\end{equation}
The distance modulus that can be obtained by combining the different cosmological distance definitions is given by \footnote{This distance modulus is given in terms of \textit{Mpc}.}
\begin{equation}
	\mu =  25-5\times\log_{10}\left[3000\bar{h}^{-1}(1+z)\int^{z}_{0}\frac{dz^{\prime}}{E(z^{\prime})}\right]\;,
\label{eq: Distance modulus}
\end{equation}
where $\bar{h} = H_0/100$. In this model, the Hubble parameter is characterise by $\alpha, \beta, \lambda, H_0, \Omega_{m}$ and $\Omega_{\eta}$. In the following two consecutive subsections, the model-free parameters  $\textbf{p}=\textbf{p}(\alpha, \beta, \lambda, H_0, \Omega_{m} )$ that will be determined from the observational data, and the detailed analysis of the late-time accelerating expansion of the universe will be studied using the constraining values of these parameters. 
 
\subsection{Constraining cosmological parameters}
\begin{figure}[ht!]
    \includegraphics[scale =0.35]{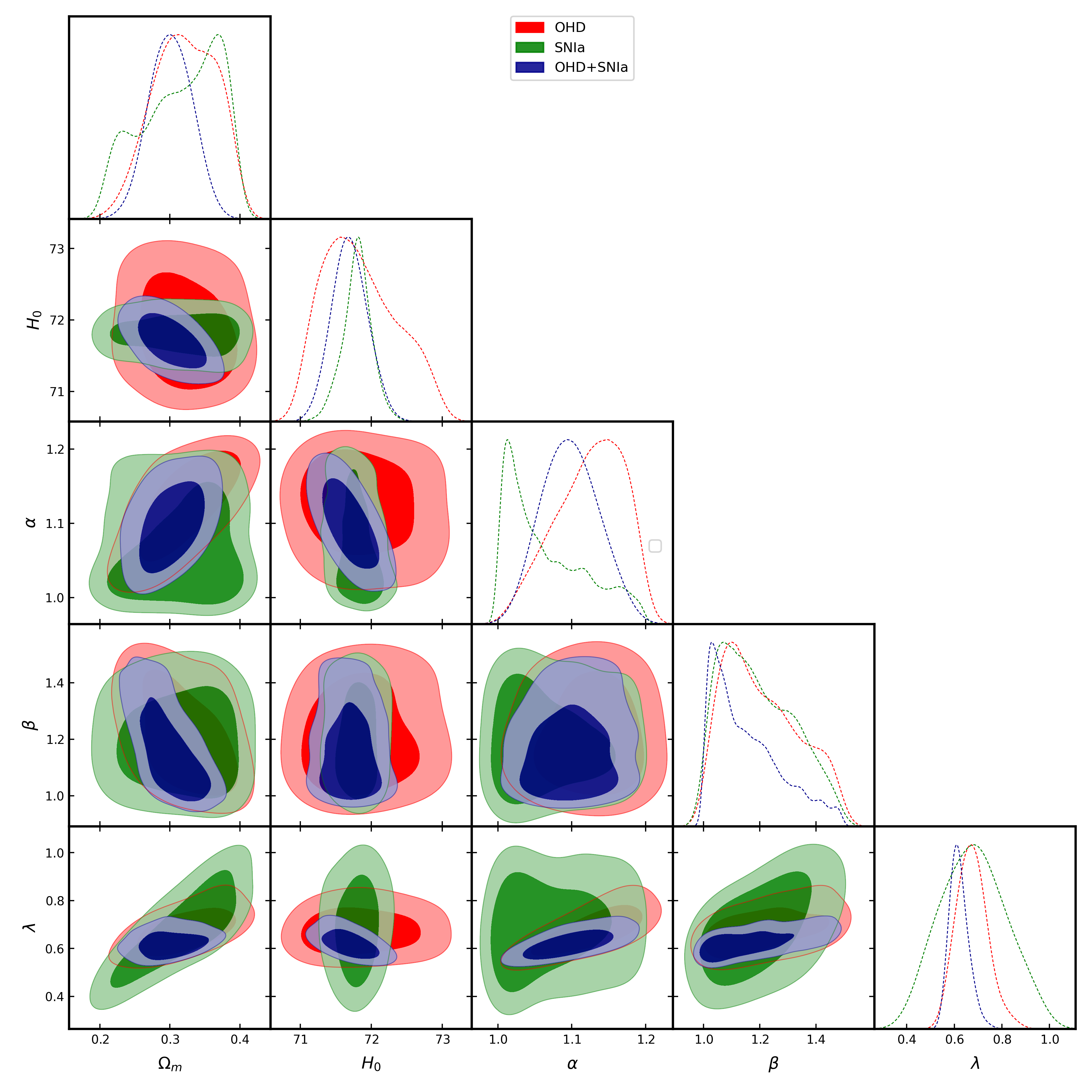}
    \caption{The best-fit values of $\Omega_m$, $H_0$, $\alpha$, $\beta$ and $\lambda$ for the $f(R,L_m)$ gravity model are shown. The values were calculated using combined datasets \texttt{OHD, SNIa,} and \texttt{OHD+SNIa} at $1\sigma$ and $2\sigma$ confidence levels.}
    \label{fig:enter-label}
\end{figure}

\begin{figure}
  \caption{Hubbel parameter for $\Lambda$CDM and $f(R,L_m)$ using the best-fit values from  \texttt{OHD}, \texttt{SNIa} and \texttt{OHD+SNIa}}
  \label{fig:enter-labelH_z1}
 	\hfill
 	\begin{minipage}{0.5\textwidth}
 		\includegraphics[scale=0.55]{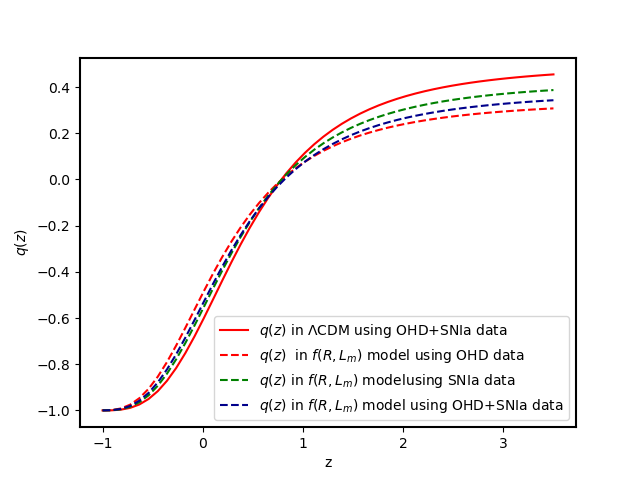}
	\end{minipage}
  \caption{Deceleration parameter, $q(z)$ is shown  for $\Lambda$CDM and $f(R,L_m)$. We use the best-fit values of the constraining parameters $Omega_m$, $H_0$, $\alpha$, $\beta$  and $\lambda$ using from  \texttt{OHD}, \texttt{SNIa} and \texttt{OHD+SNIa} for illustrative purpose. see Table \ref{Tableone} }
   \label{fig:enter-labelDecc}
  \end{figure}

\begin{figure}
 	\begin{minipage}{0.49\textwidth}
 		\includegraphics[scale=0.53]{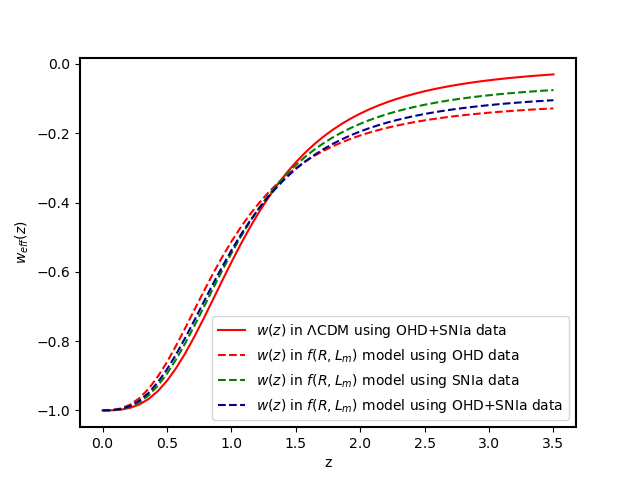}
	\end{minipage}
   \caption{Equation of State parameter $w_{eff}(z)$ versus cosmological red-shift for \lcdm and $f(R,L_m)$ using the best-fit values from Table \ref{Tableone} using  \texttt{OHD}, \texttt{SNIa} and \texttt{OHD+SNIa} data.}
    \label{fig:enter-labeldensity}
    \end{figure}

\begin{figure}
\begin{minipage}{0.49\textwidth}
     \includegraphics[scale = 0.55]{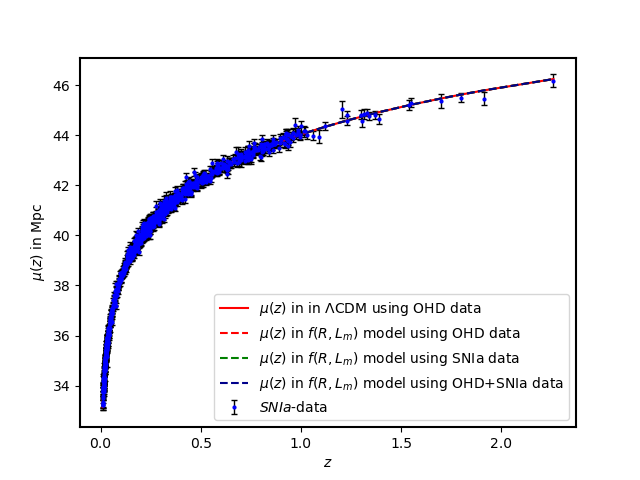}
    \caption{The distance modules ($\mu(z)$) diagram with cosmological redshift.  For illustrative purpose, we use constraining parameters $Omega_m$, $H_0$, $\alpha$, $\beta$  and $\lambda$ using from Table \ref{Tableone} for  \texttt{OHD}, \texttt{SNIa} and \texttt{OHD+SNIa}.}
    \label{fig:enter-labelmu} 
\end{minipage}
\end{figure}

In this section,  the constraining of the cosmological model parameters, $\{\Omega_m, H_0, \alpha, \beta, \lambda\}$ as presented in Table. \ref{Tableone}. Both the MCMC hammer \cite{foreman2013emcee} and GetDist \cite{lewis2019getdist} software packages have been implemented for parameter estimations, and the combined MCMC results using \texttt{OHD, SNIa} and \texttt{OHD+SNIa} have been presented in  Fig. \ref{fig:enter-label} for $f(R, L_m)$-gravity. {Using the best-fit values of the constraining parameters from Table \ref{Tableone}, the numerical results of the following parameters are presented: i) The Hubble parameter \( H(z) \) in Fig \ref{fig:enter-labelH_z1} provides insights into the expansion rate at different epochs, differentiates between cosmological models, estimates the age of the universe, and aids in understanding the universe's composition. ii) The deceleration parameter \( q(z) \) in Fig. \ref{fig:enter-labelDecc} shows whether the universe is accelerating or decelerating, with positive values indicating deceleration and negative values indicating acceleration. This parameter helps identify the transition from a matter-dominated deceleration to a dark energy-dominated acceleration, offering clues about the nature of dark energy. iii) The equation of state parameter \( w(z) \) in Fig. \ref{fig:enter-labeldensity} describes the relationship between pressure \( p \) and density \( \rho \) of the universe's components, especially dark energy, allowing cosmologists to predict the future expansion behavior of the universe. iv) The distance modulus \( \mu(z) \) in Fig. \ref{fig:enter-labelmu} is crucial in the cosmic distance ladder, facilitating measurements of cosmological distances, constraining cosmological parameters and models, and exploring the universe's geometry and curvature. These results are presented for both \(\Lambda\)CDM and \( f(R, L_m) \)-gravity theories. Together, these parameters \(H(z), q(z), w_{eff}(z), \mu(z)\) provide a comprehensive understanding of the universe's expansion history, its components, and the underlying cosmological models.}

{
\begin{table}[ht!]
    \begin{tabular}{ |c|c|c|c|c|}
    \hline
Dataset &Parameters & $\Lambda$CDM & $f(R,L_m)$ gravity \\
	\hline
 \texttt{OHD}& $\alpha$ & - &   $1.130^{+0.057}_{-0.026}$\\
 &	$\beta$ & -& $1.26^{+ 0.210}_{-0.110}$  \\
 &	$\lambda$ &- &  $0.691^{+0.072}_{-0.084}$ \\
&	$\Omega_{m}$ &$0.259^{0.028}_{0.038}$ &  $0.328^{+0.052}_{-0.033}$ \\
&	$H_{0}$  &  $70.79^{+1.22}_{-1.21}$& $71.99_{-0.74}^{+0.45}$  \\
	\hline
 \texttt{SNIa} & $\alpha$ & - & $1.052^{0.023}_{0.053}$  \\
   &	$\beta$ &- & $1.162^{+ 0.060}_{-0.16}$  \\
 &	$\lambda$ & -&  $0.640^{+0.15}_{-0.17}$ \\
&	$\Omega_{m}$ & $0.278^{0.025}_{0.024}$ &  $0.306^{+0.058}_{-0.058}$ \\
&	$H_{0}$  & $72.78^{+0.47}_{-0.47}$ & $71.82^{+0.23}_{-0.23}$  \\
	\hline
 \texttt{OHD+SNIa}& $\alpha$ & - &$1.091^{+0.035}_{0.042}$\\
   &	$\beta$ & -& $1.237^{+ 0.056}_{-0.16}$  \\
 &	$\lambda$ &- &  $0.630^{+0.031}_{-0.050}$ \\
&	$\Omega_{m}$ & $0.261 ^{0.019}_{0.013}$&  $0.287^{+0.031}_{-0.031}$\\
&	$H_{0}$  & $72.192_{-0.182}^{0.195}$ & $71.72_{-0.23}^{+0.26}$  \\
 \hline
\end{tabular}
    \caption{The best-fit values of parameters $\Omega_m, H_0, \alpha, \beta, \lambda$ are determined for both models, $\Lambda$CDM and $f(R,L_m)$, using three data sets: \texttt{OHD, SNIa} and \texttt{OHD+SNIa}. From the full-mission Planck measurements of the cosmic microwave background (CMB) \cite{aghanim2020planck}, the values of the Hubble constant $H_0 = (67.4 \pm 0.5)$ in ${km s}^{-1}{Mpc}^{-1}$ and the matter density parameter $\Omega_m = 0.315 \pm 0.007$ are obtained.
    }
    \label{Tableone}
\end{table}
}
\begin{figure}[ht!]
 	\begin{minipage}{0.49\textwidth}
 		\includegraphics[scale = 0.53]{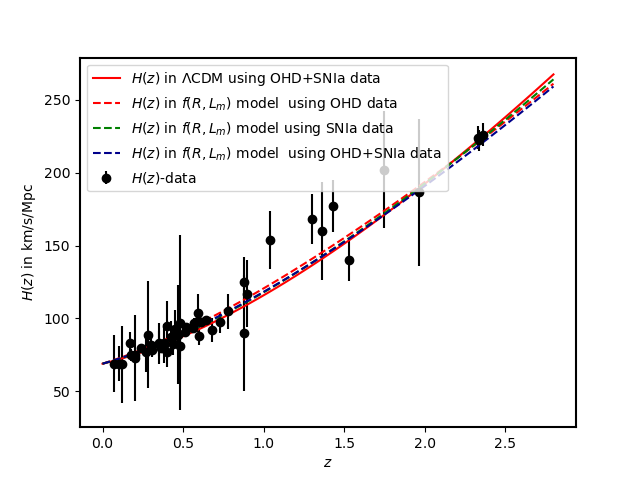}	
 	\end{minipage}
  \end{figure}
\subsection{Statistical analysis}
We conduct a statistical analysis using the Akaike Information Criterion (AIC) and the Bayesian/Schwarz Information Criterion (BIC) methods to evaluate the viability of the \( f(R, L_m) \) gravity models in comparison to the \(\Lambda\)CDM model. For the sake of comparison, we treat the \(\Lambda\)CDM model as the "accepted" benchmark to validate the \( f(R, L_m) \) gravity model based on the AIC and BIC criteria. These criteria allow us to establish the acceptance or rejection of the $f(R,L_m)$-gravity model. The $AIC$ and $BIC$ values in the $\Lambda$CDM and $f(R,L_m)$ gravity models are calculated considering the following formulation:
\begin{equation}
    \begin{split}
        \bullet\quad AIC &= \chi ^{2} +2K,\\
        \bullet\quad BIC &= \chi ^{2} +K\log(N_i),
    \end{split}
\end{equation}
where $\chi^{2}$ is calculated using the model's Gaussian likelihood function $\mathcal{L}(\hat{\theta} |data)$ value and $K$ is the number of free parameters for the particular model. At the same time, $N_i$ is the number of data points for the $i^{th}$ data set. In this context, we will use Jeffrey's scale to quantify the degree to which the \( f(R, L_m) \) gravity model should be "accepted" or "rejected" compared to \(\Lambda\)CDM. Specifically, a \(\Delta IC \leq 2\) indicates that the proposed theoretical model has substantial observational support for the fitted data\footnote{\textbf{N.B} \(\Delta IC\) represents both \(\Delta AIC\) and \(\Delta BIC\)}, a \(4 \leq \Delta IC \leq 7\) suggests less observational support and a \(\Delta IC \geq 10\) signifies no observational support.
\begin{table*}
\caption{The calculated statistical values of  ${\mathcal{L}(\hat{\theta}|data)}$,  $\chi ^{2}$, $\chi^{2}_{\nu}$, AIC, $|\Delta AIC|$, BIC and  $|\Delta BIC|$  for both models ($\Lambda$CDM) and $f(R,L_m)$-gravity) using \texttt{OHD}, \texttt{SNIa} and \texttt{OHD}+\texttt{SNIa} datasets.}
\label{sphericcase}
\begin{tabular*}{\textwidth}{@{\extracolsep{\fill}}lrrrrrrrl@{}}
\hline
\textbf{Data }& Model & \textbf{$\mathcal{L}(\hat{\theta}|data)$} & \textbf{$\chi ^{2}$} &$\chi^{2}_{\nu}$& \textbf{$AIC$} &\textbf{$|\Delta AIC|$} & \textbf{$BIC$} & \textbf{$|\Delta BIC|$} \\
\hline
  &$\Lambda CDM$&&&&&&&\\
   \hline
		\texttt{OHD}  & & -16.1 & 33.9 & 0.70 &39.9 & 0 & 45.7 & 0\\
  \texttt{SNIa} && -517.8& 1035.7  &  0.99 & 1039.7 & 0 & 1049.6 & 0\\
   \texttt{OHD+SNIa} && -537.96& 1075.9  &  0.85 & 1069.2 & 0 & 1091.3 & 0\\
			\hline
    & $f(R,L_m)$& &&& & &&\\
   \hline
   \texttt{OHD} &  &-14.4 & 28.8 & 0.6 & 38.8 & 1.1 & 48.5& 2.8\\
   \texttt{SNIa} & &  -518.7 & 1037.4 & 0.99 & 1047.4 & 7.7 & 1072.3&22.7 \\
   \texttt{OHD}
   +\texttt{SNIa} & &  -531.8 &1063.6 & 0.97 & 1073.1 & 3.8 & 1098.01& 6.7\\
			\hline
\end{tabular*}
\end{table*}
As presented in Table \ref{Tableone}, the best-fit parameters of $\Omega_m$, $H_0$, $\alpha$, $\beta$ and $\lambda$ are constrained using the \texttt{OHD}, \texttt{SNIa}, and \texttt{OHD + SNIa} datasets. 
\\
\\
Then, the total $\chi^2$ and other statistical quantities, namely the likelihood function ${\mathcal{L}(\hat{\theta}|data)}$, the reduced Chi-Square,  $\chi^{2}_{\nu}$, AIC, the change of  AIC, $|\Delta AIC|$, BIC and  the change of BIC, $|\Delta BIC|$ are presented in Table \ref{sphericcase}  for \texttt{OHD, SNIa}, and \texttt{OHD + SNIa}.  Our statistical results show that the $\Delta AIC$ values for the $f(R, L_m)$ gravity model are $1.1$, $7.7$, and $3.8$ for the \texttt{OHD}, \texttt{SNIa}, and \texttt{OHD + SNIa} datasets, respectively. These results indicate substantial observational support for the model with the \texttt{OHD} and \texttt{OHD+SNIa} data but not when using \texttt{SNIa}. And the corresponding values of the $\Delta BIC$ values for the same model are $2.8$, $22.7$, and $6.7$ for the respective datasets, suggesting that the $f(R, L_m)$ gravity model lacks observational support for the \texttt{SNIa} data and has less support for the combined \texttt{OHD + SNIa} data based on Jeffrey's scale criteria. Due to this discrepancy in the competitiveness of the $f(R, L_m)$ model, we are compelled consider the linear cosmological perturbations and conduct further investigations to compare the predictions of the $f(R, L_m)$ gravity model with relevant growth structure data $f$ and $f\sigma_8$.
\section{Cosmological perturbation equations}\label{pert}
One way of checking the viability of cosmological and gravitational models is by scrutinising their predictions of large-scale structure formation. This is often done using the powerful tool of perturbation theory, which treats the real universe as lumpy and full of inhomogeneities that lead to the formation of galaxies, clusters, voids, filaments, and walls. According to cosmological perturbation theory, the seeds of such structures, set up during inflation, were amplified due to gravitational instabilities \cite{lifshitz1946gravitational} in the expanding universe. While there are two ways of studying the mechanism of these gravitational instabilities and how the seeds evolve in an expanding background, the metric \cite{bardeen1980gauge, kodama1984cosmological} and the covariant \cite{ehlers1961beitrage,hawking1966perturbations, olson1976density} approaches, we will follow the latter in this work. While the two approaches have their own pros and cons, one can in principle write down a correspondence between them once a specific gauge is chosen (see~\cite{abebe2012covariant} and the references therein for more details). 
\\
\\{
In this section, the growth structure of the universe whose underlying gravitational theory is the $f(R, L_m)$ gravity model is studied using the $1+3$ covariant and gauge-invariant formalism of perturbations. We do this by defining the covariant and gauge-invariant gradients of the energy density \(\rho_m\) and the volume expansion \(\theta =3H\), respectively, as:
\begin{eqnarray}
	&&\Delta^m_a=\frac{a}{\rho_m}\tilde{\nabla}_a\rho_m\;, \qquad 
	Z_a=a\tilde{\nabla}_a\theta\;.\label{36}
\end{eqnarray}
Here the differential operator $\tilde{\nabla}_a$ defines the covariant spatial derivative.
The linearised evolution of the volume expansion is given by the Raychaudhuri equation
\begin{eqnarray}
	&&\dot{\theta}=-\frac{1}{3}\theta^2-\frac{1}{2}(1+3w)\rho +\nabla^a\dot{u}_a \;,
\end{eqnarray}
where $u_a$ is the four-vector velocity of the matter fluid and $w$ is the equation of state parameter for the matter fluid. The corresponding conservation equations for the fluid (assumed perfect) are given by
\begin{eqnarray}
	&&\dot{\rho}+\theta(\rho+p) = 0\;,\\&&
	(\rho+p)\dot{u}_a+\tilde{\nabla}_a p = 0\;.
\end{eqnarray} 
Taking the time derivatives of the gauge-invariant variables Eqs. \eqref{36}, we obtain the following system of first-order evolution equations:
\begin{eqnarray}
	&& \dot{\Delta}^m_a=-(1+w)Z_a+w\theta \Delta^m_a \;,\label{density} \\
	&&\dot{Z}_a=- \frac{2\theta}{3}Z_a-\left(\frac{\rho}{2} + \frac{w\tilde{\nabla}^2}{1+w}\right) \Delta^m_a\;.\label{volume}
\end{eqnarray}
As presented in \cite{sahlu2020scalar,abebe2012covariant,sami2021covariant,Ntahompagaze}, scalar perturbations play a significant role in the formation of large-scale structures. We introduce the method of isolating any scalar variable $Y$ from the first-order evolution equations. This is achieved through the standard decomposition, resulting in:
	\begin{equation}
		a\nabla_aY_b=Y_{ab}=\frac{1}{3}h_{ab}Y+\Sigma_{ab}^Y+Y_{[ab]}\;,
	\end{equation}
where $Y=a\nabla_a Y^a$, whereas $\Sigma^Y_{ab}=Y_{(ab)}-\frac{1}{3}h_{ab}Y$ and $Y_{[ab]}$ 
represent the shear (distortion) and vorticity (rotation) of the density gradient field, respectively. We applied the same decomposition techniques using the scalar perturbation equations derived from Eqs. \eqref{density} - \eqref{volume}, resulting in:
\begin{equation}
		{\delta}_m=a\tilde{\nabla}^a\Delta^m_a,\hspace{0.5cm}	\mathcal{Z}=a\tilde{\nabla}^a Z_a\;.
	\end{equation}
These variables evolve according to the first-order perturbation evolution equations:
\begin{eqnarray}
	&& \dot{\delta}^m_a=-(1+w)\mathcal{Z}_a+w\theta \delta^m_a \;,\label{density1} \\
	&&\dot{\mathcal{Z}}_a=- \frac{2\theta}{3}\mathcal{Z}_a-\left(\frac{\rho}{2} + \frac{w\tilde{\nabla}^2}{1+w}\right) \delta_a\label{volume1}
\end{eqnarray}
After identifying the system \eqref{density1} - \eqref{volume1} of scalar evolution equations, we apply the harmonic decomposition method as outlined in \cite{abebe2012covariant, ntahompagaze2018study, sami2021covariant} to obtain the eigenfunctions and the corresponding wave number $\tilde{\nabla}^2  \equiv -{k^2}/{a^2}$ (where the wave number $k = \frac{2\pi a}{\lambda}$ \cite{dunsby1992cosmological} and $\lambda$ is the wavelength) for harmonic oscillator differential equations in $f(R,L_m)$ gravity. To extract eigenfunctions and wave numbers, the harmonic decomposition technique is applied to the first-order linear cosmological perturbation equations of scalar variables \cite{sahlu2020scalar} as those in \eqref{density1} - \eqref{volume1}. For any second-order functions $X$ and $Y$ the harmonic oscillator equation is given as
	\begin{equation}
		\ddot{X}=A\dot{X}+BX-C(Y,\dot{Y} ),
	\end{equation}	
	where the frictional force, restoring force, and source force are expressed by $A$, $B$, and $C$, respectively, and the separation of variables takes the form
		$X=\sum_{k}X^k(t)Q^k(x), \hspace{0.1cm}{\rm and}  \hspace{0.1cm}	Y=\sum_{k}Y^k(t)Q^k(x)\;,$
	where $k$ is the wave number and $Q^k(x)$ is the eigenfunction of the covariantly defined Laplace-Beltrami operator in (almost) FLRW space-times, $
		\nabla^2Q^k(x)=-\frac{k^2}{a^2}Q^k(x).$
After we perform the scalar and harmonic decomposition techniques,  the second-order evolution equation is yielded as
 \begin{eqnarray}
		&&\ddot{{\delta}}^k= -\left( \frac{2\theta}{3}-w\theta\right)\dot{{\delta}}^k\nonumber\\&&-\Bigg[\frac{\rho}{2}(1-2w-3w^2)-\frac{k^2w}{a^2}\Bigg]{\delta}^k\;.\label{secondo}
	\end{eqnarray}
In this manuscript, we assume a matter-dominated universe with $w=0$, considering that the formation and evolution of large-scale structures, galaxy clusters, and voids significantly depend on the matter components of the universe. Matter's gravitational influence is crucial during the initial stages of structure formation and in determining the observable characteristics and intricate details of galaxies and galaxy clusters. We shall  also use the redshift-space transformation technique so that any first-order and second-order time derivative functions $\dot{Y}$ and $\ddot{Y}$ become
\begin{eqnarray}\label{transformation}
   &&\dot{Y} = -(1+z)HY'\;,\nonumber\\&&
 \ddot{Y} = (1+z)H^2Y' +(1+z)^2H^2Y''+  (1+z)^2H'H Y'\;.
\end{eqnarray}
By admitting Eq. \eqref{transformation}, the second-order time derivative of the evolution equation Eq. \ref{secondo} read as
  \begin{eqnarray}\label{secondo1}
   &&\delta''_m= \left( 1-\frac{1}{(1+z)}\frac{H'}{H}\right){{\delta}}'_m- \frac{3}{2}\mathcal{R}(z){\delta}_m
 \end{eqnarray}
 since $\mathcal{R}(z) \equiv \frac{\Omega_{m} (1+z)^{\frac{\alpha+1}{2\alpha-1}}}{E^2(z)}\;.$
\section{Structure growth in $f(R,L_m)$ gravity}\label{growth}
The growth structure of the universe is influenced by dark matter, dark energy, and the initial conditions set by the Big Bang. Observing the distribution of galaxies aids in understanding the underlying cosmological model, as the rate at which cosmic structures grow provides clues about the nature of dark energy. Faster growth rates suggest a universe with less influence from dark energy, while slower rates indicate stronger dark energy effects, leading to accelerated cosmic expansion. Understanding growth structure refines models of the universe's origin and fate, offering insights into its age, composition, and ultimate destiny. Large-scale surveys, such as the Sloan Digital Sky Survey, map these cosmic structures to study growth patterns, and analyzing cosmic microwave background radiation provides information about early growth structures. These observations connect to theories of cosmic inflation and the formation of the cosmic web, helping to test and constrain various cosmological theories and parameters.
\\
\\
One of the main aspects of this work is the study of the growth structure of the universe within the framework of the $f(R, L_m)$-gravity model by implementing the $1+3$ covariant formalism. In this section, we examine the ability of the $f(R, L_m)$-gravity model to statistically fit large-scale structure data. The large scale structure data used for comparison with the observational data are sourced from Tables \ref{datasetsofgrowth}, providing relevant empirical information on the growth of cosmic structures. As presented in \cite{linder2003cosmic,wang1998cluster,avila2022inferring}, the growth factor is a crucial measure that describes how cosmic structures, such as galaxies and clusters of galaxies, evolve over time due to gravitational instability. The growth factor denoted by $D(z)$ 
\begin{eqnarray}
     D(z) = \frac{\delta(z)}{\delta(z=0)}
 \end{eqnarray}
is a function of the $z$ representing the expansion of the universe. The growth factor quantifies the amplitude of density perturbations at any given time relative to their initial values. It is often normalized to be $\delta(z_{in}) = 1$ at the present time ($z =0$). Mathematically, the growth factor's evolution is governed by a differential equation that includes the Hubble parameter and the matter density parameter \cite{lee2014measuring}. This factor is essential for understanding how small initial over-densities in the matter distribution grow due to gravitational attraction, leading to the formation of large-scale structures. The growth factor's dependence on the universe's composition, including dark matter and dark energy, highlights how these components influence structure formation. In a dark energy-dominated universe, the growth of structures slows as the accelerated expansion counteracts gravitational collapse. The growth factor is vital for modeling galaxy formation and large-scale structures, and for comparing theoretical predictions with observations from the cosmic microwave background (CMB), galaxy surveys, and other large-scale structure surveys. Additionally, the related growth rate $f(z)$, which measures the rate at which structures grow, is used in various observational probes, including redshift-space distortions. The growth rate $f(z)$, as obtained from the density contrast $\mathcal{\delta}_m$, yields 
\begin{equation}
    f 
    \equiv       
    \frac{{\rm d}\ln{{{\delta}}_m}}{{\rm d}\ln{a}} = -(1+z)\frac{\delta'_m(z)}{\delta_m(z)}    
    \;.\label{growth1}
\end{equation}
Thus, the growth factor is fundamental in understanding the dynamic evolution of the universe's structure. Thus, by substituting the definition of \eqref{growth1} into the second-order evolution equation \eqref{secondo1}, the evolution of the growth rate is governed by the following expression\footnote{For the case of $\Lambda$CDM, it is straightforward to obtain
\begin{eqnarray}\label{growthratelcdm}
    && (1+z)f' = f^2 -\left[(1+z)\frac{H'}{H}-2\right] f -\frac{3  \Omega_m}{2E^2}(1+z)^3 \label{rr1}\;. \nonumber\\&&
\end{eqnarray}}
\begin{eqnarray}\label{growthrate}
     (1+z)f' &=& f^2 -\left[ (1+z)\frac{H'}{H} -2\right]f -\frac{3}{2}\mathcal{R}(z)\label{rr1}\;.
\end{eqnarray}  
As presented in \cite{wang1998cluster,avila2022inferring} a good approximation of the growth rate $f(z)$ is yield as\\
\begin{eqnarray}
    f(z) = \tilde{\Omega}^\gamma_m(z)\;,
\end{eqnarray}
where $\tilde{\Omega}(z) = \frac{\Omega_m(z)}{H^2(z)/H^2_0}$, and $\gamma$ is the growth index. As presented in \cite{linder2007parameterized}, the theoretical values growth index values,  $\gamma = 3(w-1)/(6w-5)$. For the case of $w = -1$, in the $\Lambda$CDM model the value of $\gamma = 6/11$, but this value varies for different alternative gravity models.
\subsection{Constraining cosmological parameters}
By implementing the MCMC simulations, the constrained parameters are provided in  Figs. \ref{fig:enter-labellcdmf}  using \texttt{f} datasets. The predicted value of \(\gamma\) is \(0.549 \pm 0.029\) using \texttt{f}-data in the $\Lambda$CDM model, while in the $f(R, L_m)$ analysis, \(\gamma\) is \(0.555\pm {0.014}\), as shown in Fig. \ref{fig:enter-labellcdmf}. Additionally, the values of \(\gamma\) and the parameters \(\{\Omega_m, \alpha, \beta, \lambda\}\) are constrained using \texttt{f} datasets with the results detailed in Table \ref{growthdata}.
\begin{table}[ht!]
    \begin{tabular}{cccc}
   \hline
Dataset &Parameters & $\Lambda$CDM & $f(R,L_m)$ gravity \\
	\hline
 \texttt{f}& $\alpha$ & - &   $1.15^{+0.020}_{-0.020}$\\
 &	$\beta$ & -& $1.12^{+ 0.13}_{-0.30}$  \\
 &	$\lambda$ &- &  $0.72^{+0.21}_{-0.13}$ \\
&	$\Omega_{m}$ & $0.251^{0.049}_{0.049}$&  $0.242^{0.016}_{0.032}$\\
&	$\gamma$  & $0.561_{-0.011}^{+0.067}$  & $0.555_{-0.014}^{+0.014}$  \\
	\hline
 \texttt{$f\sigma_8$} & $\alpha$ & - & $0.767^{0.027}_{0.064}$  \\
   &	$\beta$ &- & $1.08^{+ 0.40}_{-0.16}$  \\
 &	$\lambda$ & -&  $0.281^{+0.075}_{-0.110}$ \\
&	$\Omega_{m}$ & $0.278^{0.025}_{0.024}$ &  $0.283^{+0.036}_{-0.048}$ \\
&$\sigma_8$  & $0.805^{+0.044}_{-0.081}$& $0.798^{+0.045}_{0.086}$  \\
 \hline
\end{tabular}
    \caption{The best-fit values of parameters $\Omega_m$, $\alpha$, $\beta$, $\lambda$, and $\gamma$ are constrained for both models, $\Lambda$CDM and $f(R,L_m)$, using three data sets: \texttt{f} and \texttt{f$\sigma_8$}. From the full-mission Planck measurements of the cosmic microwave background (CMB) \cite{aghanim2020planck}, the value for the matter fluctuation amplitude ($\sigma_8$) is obtained as $\sigma_8 = 0.811\pm 0.006$.
    }
    \label{growthdata}
\end{table}
From these plots, we notice that the constraining parameters
$\Omega_m = 0.242^{+0.016}_{-0.032}$, $\alpha = 1.15^{+0.20}_{-0.20}$, $\beta = 1.12^{+0.13}_{-0.30}$, $\lambda = 0.72^{+0.30}_{-0.13}$  and \(\gamma = 0.555\pm {0.014}\) at $1\sigma$ and $2\sigma$ confidence levels. We also presented the numerical solutions of the growth factor $D(z)$ as shown in Fig. \ref{fig:enter-labelgrowthfactor}
and the evolution of growth rate $f(z)$ diagram as shown in Fig. \ref{fig:enter-labelgrowthrate}.  
 \begin{figure}
 \begin{minipage}{0.4\textwidth}
 		\includegraphics[scale=0.8]{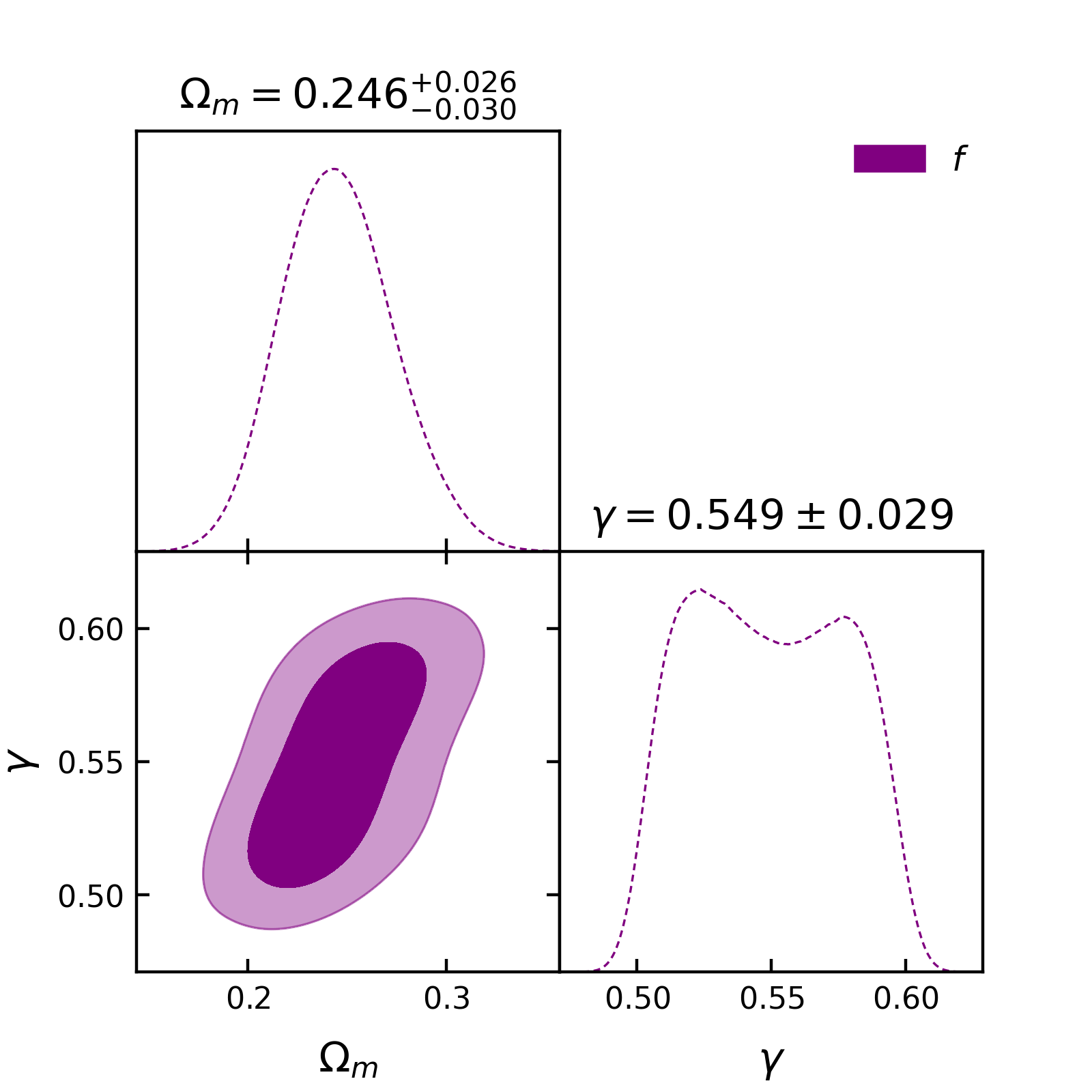}
	\end{minipage}
 \caption{The constrained parameters of $\Omega_m = 0.246^{+0.026}_{-0.030}$, $\gamma = 0.549\pm0.029$ are shown in $\Lambda$CDM model  using \texttt{f}-data.}
    \label{fig:enter-labellcdmf}
     \end{figure}
    \begin{figure}
 \begin{minipage}{0.5\textwidth}
 		\includegraphics[scale=0.36]{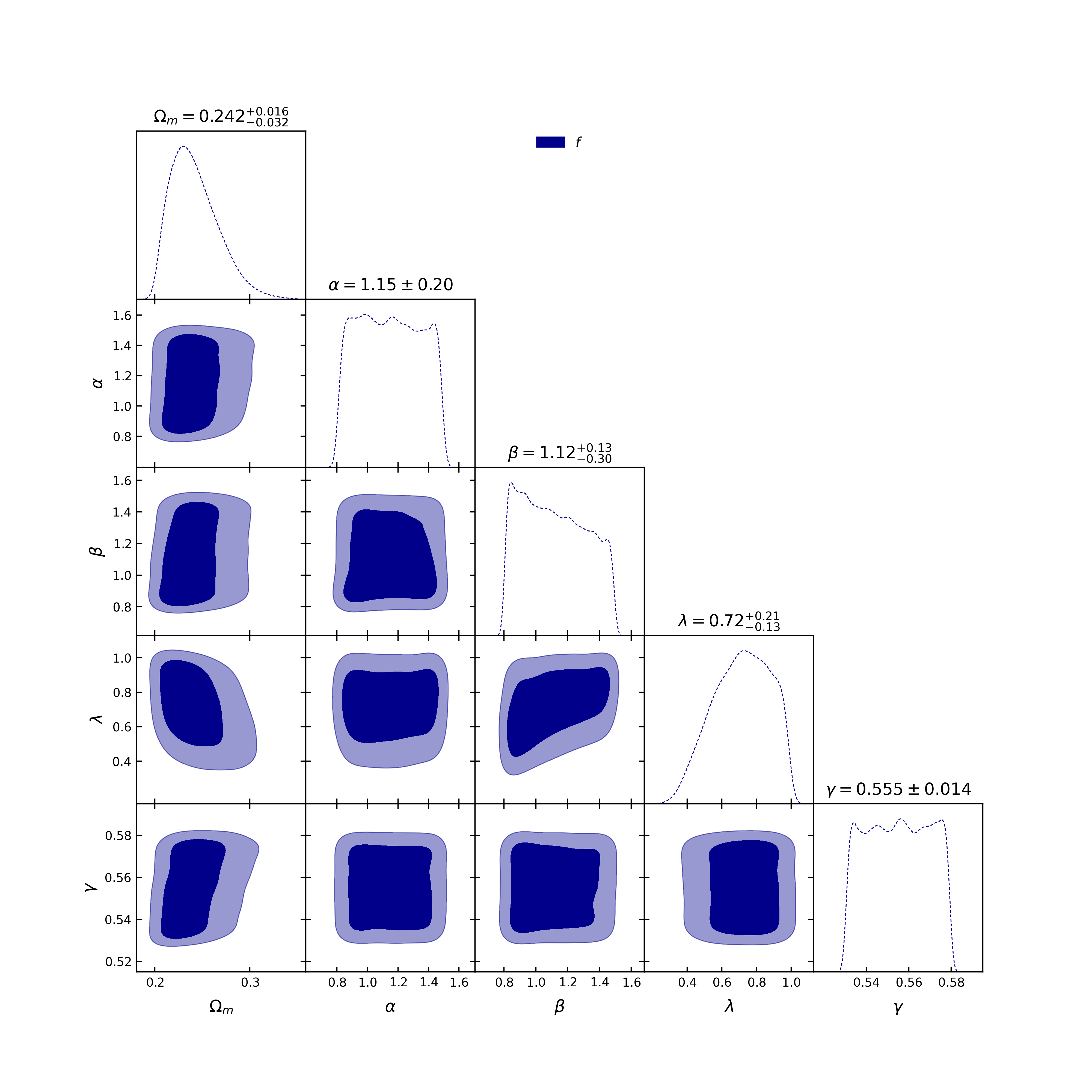}
	\end{minipage}
   \caption{ The constrained parameters of $\Omega_m = 0.242^{+0.016}_{-0.032}$, $\alpha = 1.15^{+0.20}_{-0.20}$, $\beta = 1.12^{+0.13}_{-30}$, and $\lambda = 0.72^{+0.30}_{-0.13}$ are shown in $f(R,L_m)$ model using \texttt{f}-data. }
    \label{fig:enter-labellxx}   
\end{figure}
\begin{figure}
    \begin{minipage}{0.49\textwidth}
 		\includegraphics[scale=0.38]{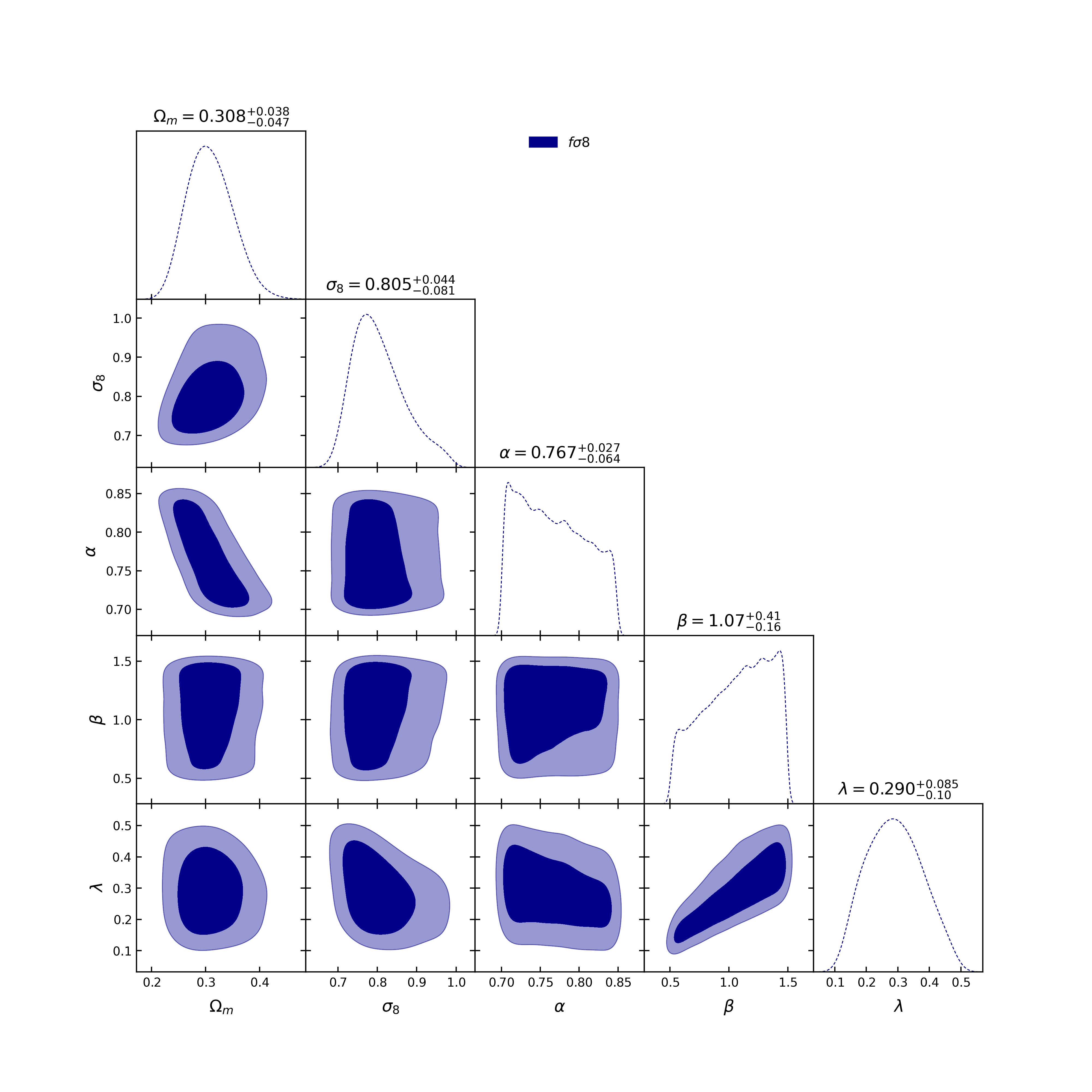}
	\end{minipage}
   \caption{The constrained parameters of  $\Omega_m = 0.284^{+0.035}_{-0.049}$, $\sigma_8 = 0.799^{+0.045}_{-0.086}$, $\alpha = 0.766^{+0.026}_{-0.064}$, $\beta = 1.08^{+0.40}_{-0.16}$, and $\lambda = 0.279^{+0.078}_{-0.11}$ are presented in $f(R,L_m)$ model using \texttt{f$\sigma_8$}-data at $1\sigma$ and $2\sigma$ confidence levels.}
    \label{fig:enter-label??}
    \label{fig:enter-labelmcmcfsigma8}
\end{figure}
A combination of the linear growth rate $f(z)$ with the root mean square normalization of the matter power spectrum $\sigma_8$ within the radius sphere $8h^{-1}$Mpc, yields the redshift-space distortion $f\sigma_8$  which directly measures the matter density perturbation rate as expressed 
\begin{eqnarray}\label{growth11}
  f\sigma_8(z)  = -(1+z)\sigma_{8,0}\delta_m'(z)\;,
 \end{eqnarray}
and $\sigma_8(z)$  for the given redshift $z$ can be expressed as \cite{hamilton1998linear}
\begin{eqnarray}\label{sigma88}
    \sigma_8(z) = \sigma_{8,0}\delta_m(z) \;.
\end{eqnarray}
In this paper, we also study the redshift space-distortion $f\sigma_8$ in $f(R,L_m)$ gravity model which refers to the apparent distortion of galaxy positions in redshift space due to their peculiar velocities. It provides valuable information about the growth rate of cosmic structures and the underlying matter density, offering insights into the dynamics of the universe's expansion and the nature of dark matter and dark energy.
We also implemented the MCMC simulation to constrain the best-fit values of $\Omega_m = 0.284^{+0.035}_{-0.049}$, $\sigma_8 = 0.799^{+0.045}_{-0.086}$, $\alpha = 0.766^{+0.026}_{-0.064}$, $\beta = 1.08^{+0.40}_{-0.16}$, and $\lambda = 0.279^{+0.078}_{-0.11}$,  using the \texttt{f$\sigma_8$} as presented in Fig. \ref{fig:enter-labelmcmcfsigma8} in $f(R, L_m)$ gravity approach. Using these constraining parameter's values,  the evolution of \texttt{f$\sigma_8$}  is presented in $f(R, L_m)$ gravity model through cosmological red-shift (see Fig. \ref{fig:enter-labelRSD}). 
\begin{figure}[h!]
 	\begin{minipage}{0.49\textwidth}
 		\includegraphics[scale = 0.5]{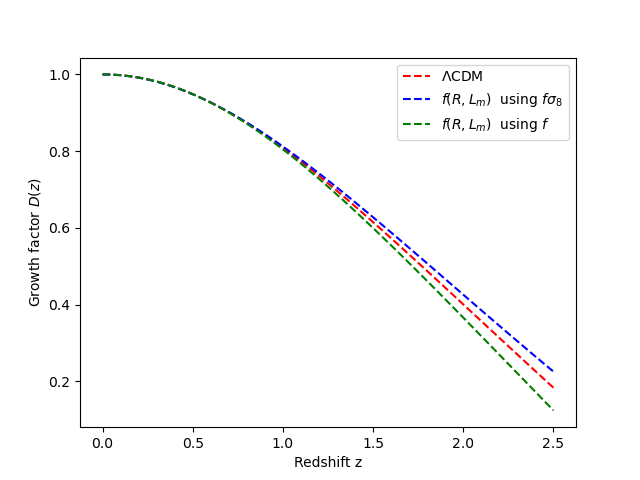}	
 	\end{minipage}
  \caption{Growth factor $D(z)$ with cosmological redshift for $\Lambda$CDM and $f(R)$-gravity model.}
  \label{fig:enter-labelgrowthfactor}
  \end{figure}
  \begin{figure}
  \begin{minipage}{0.49\textwidth}
 		\includegraphics[scale=0.53]{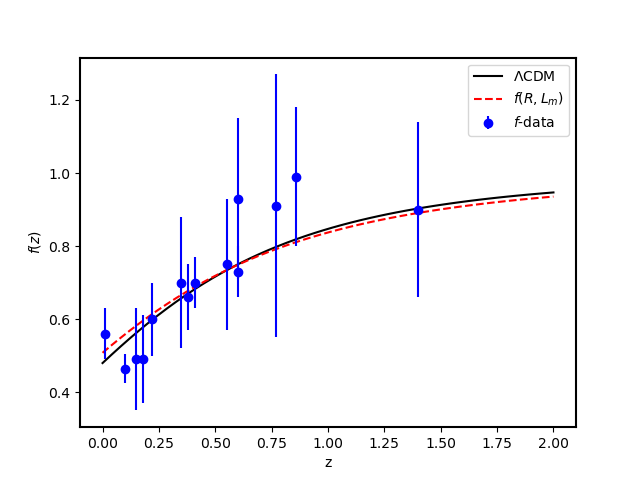}
	\end{minipage}
   \caption{The growth rate $f(z)$ with cosmological redshift in $\Lambda$CDM and $f(R,L_m)$-gravity models. The best fit values of the parameters have been considered using \texttt{f} and \texttt{f}$\sigma_8$ datasets from Table. \ref{growthdata}.}
    \label{fig:enter-labelgrowthrate}
     \end{figure}
  \begin{figure}
 	\begin{minipage}{0.5\textwidth}
 		\includegraphics[scale=0.53]{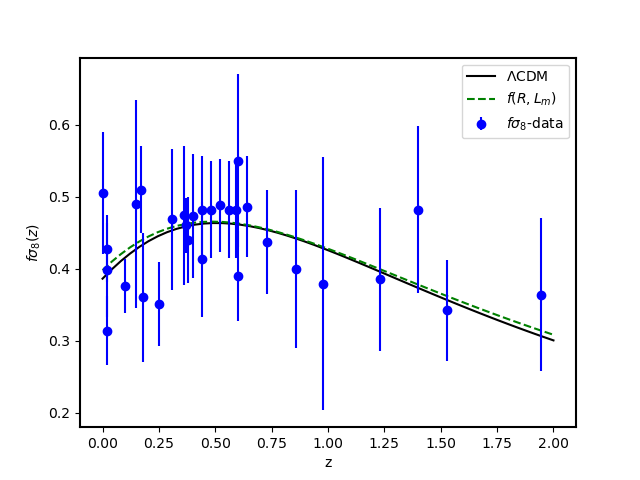}
	\end{minipage}
 \caption{The redshift space-distortion, \texttt{f}$\sigma_8$ with cosmological redshift in $\Lambda$CDM and $f(R,L_m)$-gravity models. We use the best-fit values of the parameters using \texttt{f} and \texttt{f}$\sigma_8$ datasets from Table. \ref{growthdata}.}
   \label{fig:enter-labelRSD}
\end{figure}
\subsection{The statistical analysis}
To establish the viability of the \( f(R, L_m) \) gravity model, it is essential to present and analyze several statistical values. These include the likelihood function \( \mathcal{L}(\hat{\theta}|data) \), the reduced chi-square \( \chi^{2}_{\nu} \), the chi-square \( \chi^{2} \), the Akaike Information Criterion (AIC), the absolute difference in AIC (\( |\Delta AIC| \)), the Bayesian Information Criterion (BIC), and the absolute difference in BIC (\( |\Delta BIC| \)). These statistical measures will be used to compare the performance of the \( f(R, L_m) \) gravity model against the well-established \(\Lambda\)CDM model, which will serve as the benchmark or accepted standard model in this analysis.   The statistical values will help quantify the models' goodness of fit, with lower values of AIC and BIC indicating a better model. The differences \( |\Delta AIC| \) and \( |\Delta BIC| \) will highlight the relative performance of the \( f(R, L_m) \) model compared to the \(\Lambda\)CDM model. This thorough statistical analysis is crucial in determining whether the \( f(R, L_m) \) gravity model can be considered a viable alternative to the \(\Lambda\)CDM model in explaining cosmic phenomena.
\begin{table*}
\caption{The calculated statistical values of  ${\mathcal{L}(\hat{\theta}|data)}$,  $\chi ^{2}$, $\chi^{2}$-red, AIC, $|\Delta AIC|$, BIC and  $|\Delta BIC|$  for both models ($\Lambda$CDM) and $f(R,L_m)$-gravity) using \texttt{f} and \texttt{f$\sigma_8$} datasets.}
\label{sphericcase1}
\begin{tabular*}{\textwidth}{@{\extracolsep{\fill}}lrrrrrrrl@{}}
\hline
\textbf{Data }& Model & \textbf{$\mathcal{L}(\hat{\theta}|data)$} & \textbf{$\chi ^{2}$} &$\chi^{2}_{\nu}$ & \textbf{$AIC$} &\textbf{$|\Delta AIC|$} & \textbf{$BIC$} & \textbf{$|\Delta BIC|$} \\
\hline
  &$\Lambda CDM$&&&&&&&\\
   \hline
		\texttt{f}  & & -3.126 & 6.252 & 0.521& 10.252 & 0 & 11.963 & 0\\
   \texttt{f$\sigma_8$} && -8.00& 16.00 &  0.57 & 20.00 & 0 & 24.00 & 0\\
			\hline
    & $f(R,L_m)$& &&& & &&\\
   \hline
   \texttt{f} &  &-2.01 & 4.02 & 0.630 & 14.02 & 3.767 &17.215 &5.25\\
   \texttt{f$\sigma_8$} & & -6.20 & 12.40 & 0.770 & 22.40 & 2.40 & 29.405&5.405\\
			\hline
\end{tabular*}
\end{table*}
As shown in Table. \ref{sphericcase1} the statistical values of $\Delta AIC$ for the $f(R, L_m)$ gravity model are $3.767$ and  $2.40$ comapere with the $\Lambda$CDM  using  the \texttt{f} and \texttt{f}$\sigma_8$ datasets respectively.  These statistical figures indicate that the model has substantial observational support. In contrast, the corresponding values of $\Delta BIC$ read $5.25$ and  $5.405$ for the same datasets, suggesting that the $f(R, L_m)$ gravity model has less observational support based on Jeffrey's scale criteria.
\section{Conclusion}\label{conc}
 Using recent cosmic measurements such as \texttt{OHD}, \texttt{SNIa}, and \texttt{OHD+SNIa}, along with large scale structure data like the growth rate \texttt{f} data and redshift space distortion data \texttt{f}$\sigma_8$, this paper investigated the constraints of the background and perturbed cosmology within the framework of $f(R, L_m)$ gravity, and provided a detailed analysis of the late-time accelerating universe and structure growth in it. 
 \\
 \\After we discussed the general theory of $f(R, L_m)$ gravitation and its corresponding field equations in Sec. \ref{efes}, we proposed the model $f(R, L_m) = \lambda R + \beta L_m^\alpha + \eta$}, chosen in such a way that for the case of $\alpha = \beta = 1$ and $\lambda = 1/2$, this model exactly reduces to \lcdm cosmology for the constant term $\eta=-\Lambda$. In Sec. \ref{datameth}, the general setups of the data and methodology have been shown to constrain the $f(R)$-gravity model.  {In Sec. \ref{model}, the best-fit values of the model parameters $\Omega_m$, $H_0$, $\alpha$, $\beta$, and $\lambda$ are constraining using \texttt{OHD}, \texttt{SNIa}, and \texttt{OHD+SNIa} and presented in Table. \ref{Tableone}. {For example, from the joint analysis \texttt{OHD+SNIa} the constrained values, we obtained $\Omega_m = 0.287\pm {0.031}$, $H_0 = 71.72_{-0.23}^{+0.26}$, $\alpha = 1.091^{+0.035}_{-0.042}$, \(\beta = 1.237^{+ 0.056}_{-0.16}\) and $\lambda = 0.630^{+0.031}_{-0.050}$. Using the cosmological parameter's values presented in Table \ref{Tableone}, the detailed analysis of the accelerating expansion of the late time has been discussed, and the diagram of the key cosmological parameters, $H(z)$  $q(z)$, $w(z)$ and $\mu(z)$ presented in Fig. \ref{fig:enter-labelH_z1}, \ref{fig:enter-labelDecc}, \ref{fig:enter-labeldensity} and \ref{fig:enter-labelmu} respectively. 
The corresponding statistical results have been provided and the study of $f(R,L_m)$ gravity model is, in some sense, justified based on the calculated values of ${\mathcal{L}(\hat{\theta}|data)}$,  $\chi ^{2}$, $\chi^{2}$-red, AIC, $|\Delta AIC|$, BIC, and $|\Delta BIC|$. Statistically speaking, the \( f(R, L_m) \) model has substantial support when using \texttt{OHD} data $\Delta IC \leq 2$, less observational support with the joint analysis \texttt{OHD+SNIa} $\Delta IC\leq 7$, and is rejected using \texttt{SNIa} $\Delta IC >7$ compared with $\Lambda$CDM at the background level, based on Jeffreys' scale criteria see Table \ref{sphericcase}.}
\\
\\{As discussed in Section \ref{pert}, we implemented the $1+3$ covariant gauge-invariant formalism, and the linear cosmological perturbations have been analyzed by introducing the spatial covariant gradients for the matter fluid and volume expansion. We derived the first- and second-order evolution equations using harmonic and scalar decomposition techniques, which have a significant role in studying the large-scale structure of the universe. As presented in Eq. \eqref{secondo1}, the set of density contrast equations has been derived which is crucial for understanding the formation and evolution of cosmic structures, such as galaxies, galaxy clusters, and large-scale filaments, as it quantifies the level of overdensity or underdensity in different parts of the universe. 
\\
\\
Section \ref{growth} is dedicated to examining the growth of cosmic structure within the \( f(R, L_m) \) gravity model for several reasons: i) understanding structure growth is crucial for predicting the unique patterns of formation in \( f(R, L_m) \) gravity compared to \(\Lambda\)CDM; ii) analyzing structure growth provides insights into dark energy, which affects the cosmic expansion rate and, consequently, the growth of structures; and iii) studying structure growth helps probe the universe's initial conditions, revealing details about its early history and evolution. Utilizing the large-scale structure datasets \texttt{f} and \texttt{f}\(\sigma_8\) listed in Table \ref{growthdata}, we conducted a detailed statistical analysis after determining the best-fit parameters using MCMC simulation. For instance, at the $1\sigma$ and $2\sigma$ confidence levels, respectively, the best-fit values of  $\Omega_m = 0.242^{+0.016}_{-0.032}$, $\alpha = 1.15^{+0.20}_{-0.20}$, $\beta = 1.12^{+0.13}_{-0.30}$, $\lambda = 0.72^{+0.30}_{-0.13}$  and \(\gamma = 0.555\pm {0.014}\)  using \texttt{f}-data and $\Omega_m = 0.284^{+0.035}_{-0.049}$, $\sigma_8 = 0.799^{+0.045}_{-0.086}$, $\alpha = 0.766^{+0.026}_{-0.064}$, $\beta = 1.08^{+0.40}_{-0.16}$, and $\lambda = 0.279^{+0.078}_{-0.11}$ using the \texttt{f$\sigma_8$} data in  the limit of $f(R,L_m)$ model  The most important statistical quantities, namely: ${\mathcal{L}(\hat{\theta}|data)}$,  $\chi ^{2}$, $\chi^{2}$-red, AIC, $|\Delta AIC|$, BIC, and $|\Delta BIC|$ have been calculated for both (\lcdm and $f(R,L_m)$-gravity models using the \texttt{f} and \texttt{f$\sigma_8$} datasets. Based on Jeffreys' scale criteria, the $f(R, L_m)$ gravity models show substantial observational support based on $\Delta$AIC values but less observational support based on the $\Delta$BIC values (see Table \ref{sphericcase1}). 
\\

In summary, this paper presents a thorough analysis aimed at constraining the $f(R, L_m)$ gravity model, taking into account the universe's accelerating expansion within the context of background cosmology, as well as the growth of cosmic structures at the perturbation level, supported by observational data. By constraining the free parameters \(\Omega_m, H_0\), \(\sigma_8\), \(\alpha\), \(\beta, \lambda\) at both levels, we performed a detailed statistical evaluation to assess the compatibility of the $f(R)$ gravity model against observational data, in comparison to the $\Lambda$CDM model. Our findings suggest that the $f(R, L_m)$ gravity model is significantly supported by OHD data, but shows less support when OHD+SNIa, f, and f$\sigma_8$ data are considered. Whereas the model is not supported when using SNIa datasets alone.
A similar analysis using more data, both in terms of sample size and type of data - latest or forthcoming - needs to be done before a concrete pronouncement on viability or the ruling out of the gravitational model is made. 


\newpage
\bibliographystyle{iopart-num}
\bibliography{references}
\bibliographystyle{elsarticle-harv}

\end{document}